\documentclass[aps,prd,showkeys,nofootinbib,floatfix,amsmath,amssymb,showpacs,superscriptaddress,notitlepage]{revtex4-2}

\usepackage{rotating}
\usepackage{graphicx}
\usepackage{dcolumn}
\usepackage{bm}
\usepackage{color}
\usepackage{multirow}
\usepackage{mathtools}
\usepackage[pdftex,bookmarks=true]{hyperref}
\hypersetup{
   colorlinks,
   citecolor=black,
   filecolor=black,
   linkcolor=black,
   urlcolor=black
}
\usepackage{cleveref}

\begin{document}

\title{Electromagnetic Structure of Spin-$\frac12$ Doubly Charmed Baryons in Lattice QCD}

\author{H. Bahtiyar}
\affiliation{Department of Physics, Faculty of Science and Letters, Mimar Sinan Fine Arts University, Bomonti 34380,
Istanbul, Turkey}
\date{\today}

\begin{abstract}
We compute the electromagnetic properties of spin-$\frac12$ doubly charmed baryons on 2+1 flavor lattices that have a pion mass of $\sim$ 156 MeV. The Tsukuba action is employed for the charm quark in addition to the standard isotropic Clover action to quantify the $\mathcal{O}(m_q a)$ effects. We calculate the electric and magnetic Sachs form factors and extract the magnetic moments and electric and magnetic charge radii. We also investigate the individual quark sector contributions to the charge radii and the magnetic moments. The results provide vital information to understand the size and shape of the doubly charmed baryons. We find that the two heavy charm quarks drive the charge radii and the magnetic moments to smaller values than that of light baryons. The central values of the observables that are obtained using the relativistic action for the charm quark are $5$ to $10\%$ larger than those obtained using the Clover action. Utilizing the available lattice data, we reexamine the quark mass dependence of the observables.
\end{abstract}
\pacs{14.20.Lq, 12.38.Gc, 13.40.Gp }
\keywords{charmed baryons,magnetic moment, lattice QCD}
\maketitle
\section{Introduction}
In the last decades, measurements of the charmed baryons have accelerated with the developments in experimental facilities. Many states are discovered, yet many states need to be confirmed; therefore charmed baryon sector is theoretically appealing. All of the singly charmed ground-state baryons have been experimentally observed~\cite{BEBCTSTNeutrino:1980ktj,Biagi:1983en,CLEO:1988yda,CLEO:1998wvk,Biagi:1984mu}. However, observation of doubly charmed baryons has been long overdue. In 2002, SELEX collaboration reported the doubly charmed baryon, which contains two $c$ and one $d$ quark, and the mass of the $\Xi_{cc}^{+}$ reported as $3519 \pm 1$ MeV/c$^2$~\cite{SELEX:2002wqn}. However, none of the FOCUS~\cite{Ratti:2003ez}, BABAR~\cite{BaBar:2006bab} , BELLE~\cite{Belle:2006edu}, and LHCb~\cite{LHCb:2013hvt} could confirm SELEX's result. In 2017, LHCb Collaboration discovered the isospin partner of $\Xi_{cc}^{+}$, namely $\Xi_{cc}^{++}$~\cite{LHCb:2017iph}, containing two $c$ quarks and one $u$ quark and the mass of the $\Xi_{cc}^{++}$ reported by LHCb is $3621.40 \pm 0.72 \pm 0.27 \pm 0.14$ MeV/c$^2$, around $100$ MeV larger than the SELEX result. The mass difference has been studied within various theoretical approaches~\cite{Oudichhya:2022ssc,universe:kakadiya,Soto:2021cgk,Yao:2018ifh, Lu:2017meb,Wei:2015gsa, Karliner:2014gca}.

Doubly charmed baryons sit at the top layer of spin-$\frac{1}{2}$ flavor-mixed symmetric 20-plet of the SU(4) multiplet. This layer consists of three baryons; the isospin doublets $\Xi_{cc}^{++}($ucc$)$ and $\Xi_{cc}^{+}($dcc$)$, and the isospin singlet $\Omega_{cc}$ ($scc$). Although $\Omega_{cc}$ has not been experimentally observed yet~\cite{LHCb:2021rkb}, several theoretical studies have been conducted on its mass and form factors~\cite{Shah:2016vmd,Ebert:2002ig,Roberts:2007ni,Migura:2006ep,Can:2021ehb,Can:2013tna,Ozdem:2018uue,Can:2015exa}.

Studying the electromagnetic properties of baryons provide us with valuable information about their internal structures. One can extract information about their sizes, shapes, decay widths and compare them with the experimental results. Baryons containing two charm quarks are especially exciting to explore since examining the electromagnetic properties of two heavy-quarks bound to a light-quark helps us understand the internal interaction dynamics of baryons containing heavy-quarks. In addition, the results may shed light on our understanding of the fundamental properties of QCD, such as confinement and flavor effects. Electromagnetic properties of doubly charmed baryons have been previously studied in quark models~\cite{Mutuk:2021epz,Julia-Diaz:2004yqv,Patel_2008,Albertus:2006ya,Faessler:2006ft,Jena:1986xs,Sharma:2010vv}, MIT bag model~\cite{Bernotas:2012nz,Simonis:2018rld,Zhang:2021yul}, chiral perturbation theory~\cite{HillerBlin:2018gjw,Liu:2018euh,Li:2017cfz,Li:2020uok}, QCD sum rules~\cite{Aliyev:2022rrf,Ozdem:2018uue} and lattice QCD~\cite{Can:2013zpa,Can:2013tna}.

This work focuses on calculating the electromagnetic form factors and extracting the charge radii and magnetic moments of doubly charmed baryons with near physical $u$, $d$ sea quarks in $2+1$ flavor lattice QCD. A relativistic fermion action is employed for the charm quark in addition to the standard isotropic Clover action to quantify the $\mathcal{O}(m_q a)$ effects. Since the lattice pion mass is close to the physical pion mass, we also employ the chiral extrapolation of the results calculated with Clover action using the results of Refs~\cite{Can:2013zpa, Can:2013tna}. This paper is organized as follows: Theoretical formalism, details of the lattice setup, and definitions of the form factors are given in \Cref{sec:theo}. We present our results, compare them to other works and give a discussion in \Cref{sec:res}. \Cref{sec:con} summarizes our results.
\section{Theoretical Formalism and Lattice Setup}
\label{sec:theo}
The electromagnetic current is written in the following form:
\begin{equation}
J_\mu =  \frac{2}{3} \overline{c} \gamma_\mu c + C_\ell \overline{\ell} \gamma_\mu \ell,	
\label{eq:veccurrrent}
\end{equation}
where $\ell$ denotes the flavor of the lighter quarks (u, d, and s) and $C_\ell$ represents their charge ($2/3$ or $-1/3$). We couple the current to each valence quark in the baryon allowing us to compute the electromagnetic transition form factors by evaluating the associated matrix element:
\begin{align}
\label{eq:matelC}
 \langle {\cal B}(p')|\overline c \gamma_\mu c |{\cal B}(p) \rangle  &= \overline u(p') \Big[ \gamma _\mu F_1^c(q^2) + i\frac {\sigma_{\mu \nu} q^\nu}{2 m_ {\cal B}} F_2^c(q^2) \Big] u(p),\\
 \label{eq:matelL}
 \langle {\cal B}(p')|\overline \ell \gamma_\mu \ell |{\cal B}(p) \rangle & = \overline u(p') \Big[ \gamma _\mu F_1^\ell(q^2) + i\frac {\sigma_{\mu \nu} q^\nu}{2 m_ {\cal B}} F_2^\ell(q^2) \Big] u(p),
\end{align}
where $F_1(q^2),F_2(q^2)$ are the Dirac and Pauli form factors respectively, superscripts denote the flavor of the quarks, $u(p')$ and $u(p)$ are the Dirac spinor of the baryon with the mass of $m_{\cal B}$, and $q^\mu = p^\mu - p'^\mu$ is the transferred four-momentum. combining the Eqs.\eqref{eq:veccurrrent},\eqref{eq:matelC}, and \eqref{eq:matelL}, we obtain the complete matrix element:
\begin{equation}
 \langle {\cal B}(p')|J_\mu|{\cal B}(p) \rangle  = \overline u(p') \Big[ \gamma _\mu F_1(q^2) + i\frac {\sigma_{\mu \nu} q^\nu}{2 m_ {\cal B}} F_2(q^2) \Big] u(p),
\end{equation}
where
\begin{align}
F_1(q^2) &= \frac{2}{3} F_1^c(q^2) + C_\ell F_1^\ell(q^2),\\
F_2(q^2) &= \frac{2}{3} F_2^c(q^2) + C_\ell F_2^\ell(q^2).
\end{align}
The electric and magnetic Sachs form factors of individual quark contributions are defined in terms of the Dirac and Pauli form factors as follows:
\begin{align}
\label{eq:elSquarkcont}
G_E^c(q^2)& = F_1^c(q^2) + \frac{q^2}{4m_{\cal B}^2} F_2^c(q^2),\quad G_E^\ell(q^2) = F_1^\ell(q^2) + \frac{q^2}{4m_{\cal B}^2} F_2^\ell(q^2),\\
\label{eq:magSquarkcont}
G_M^c(q^2) &= F_1^c(q^2) + F_2^c(q^2), \quad  G_M^\ell(q^2) = F_1^\ell(q^2) + F_2^\ell(q^2).
\end{align}
The total electric and magnetic Sachs form factors of the baryon become:
\begin{align}\label{elS}
G_E(q^2)& = F_1(q^2) + \frac{q^2}{4m_{\cal B}^2} F_2(q^2),\\
\label{magS}
G_M(q^2) &= F_1(q^2) + F_2(q^2).
\end{align}
The form factors are extracted using the two-point and three-point correlation functions,
\begin{align}
\label{eq:twopt}
\langle F^{{\cal B}\,{\cal B}}(t;\textbf{p};\Gamma_4)\rangle&=\sum\limits\limits_{\textbf{x}}^{ }\,e^{i\,\textbf{p}\cdot\textbf{x}} \Gamma^{\beta\,\alpha}_4\,\langle\Omega|T(\chi_{\cal B}^{\beta}(x) \overline \chi_{\cal B}^{\alpha}(0))|\Omega \rangle,\\
\langle F^{{\cal B} J_\mu {\cal B}}(t_2,t_1;{\bf p'},\textbf{p};\Gamma) \rangle&=-i\,\sum\limits\limits_{{\bf x_2},{\bf x_1}}^{ }\,e^{-i\,\textbf{p}\cdot\mathbf{x_2}}\,e^{i\,\textbf{q}\cdot\mathbf{x_1}}\Gamma^{\beta\,\alpha}\langle\Omega|T(\chi^{\beta}_{{\cal B}}(x_2) J_\mu(x_1)\overline \chi^{\alpha}_{\cal B}(0))|\Omega \rangle,
\end{align}
where $t_1$ is the time when the electromagnetic current is inserted to quark, and $t_2$ is the time when the baryon is annihilated. Gamma matrices are defined as,
  \begin{align}
 \Gamma_4 =  \frac12
  \begin{bmatrix}
    1 & 0  \\
    0 & 0 
  \end{bmatrix}, \quad \Gamma_i = \frac12 \begin{bmatrix}
    \sigma_i & 0  \\
    0 & 0 
  \end{bmatrix}. 
  \end{align}
The baryon interpolating fields are chosen as the nucleon-like form
  \begin{align}
  \chi_{\cal B}&= \varepsilon_{abc} \big(c^{T}_a (C \gamma _5) \ell_b \big)c_c,
\label{eq:interpolating}
  \end{align}
where $\ell$ denotes $s$, $u$ and $d$ quarks for $\Omega_{cc}$, $\Xi_{cc}^{++}$ and $\Xi_{cc}^{+}$ baryons respectively. The $a,b,c$ are color indices, and the charge conjugation matrix is defined as $C= \gamma_4 \gamma_2$.

In order to eliminate the normalization factors and extract the electromagnetic form factors, we define the ratio,
\begin{align}
\label{eq:Ratiospinhalf}
    R(t_2,t_1,{\bf p'},{\bf p},\Gamma,\mu) = &\frac{\langle F^{\cal B J_\mu B}(t_2,t_1;{\bf p'},{\bf p};\Gamma) \rangle }{\langle F^{\cal BB}(t_2;{\bf p'};\Gamma_4) \rangle }\times\\ 
&\sqrt{\frac{\langle F^{\cal BB}(t_2-t_1;{\bf p};\Gamma_4) \rangle \langle F^{\cal BB}(t_1;{\bf p'};\Gamma_4) \rangle \langle F^{\cal BB}(t_2;{\bf p'};\Gamma_4) \rangle }{\langle F^{\cal BB}(t_2-t_1;{\bf p'};\Gamma_4) \rangle \langle F^{\cal BB}(t_1;{\bf p};\Gamma_4) \rangle \langle F^{\cal BB}(t_2;{\bf p};\Gamma_4) \rangle }}.\nonumber
\end{align}
In the large Euclidean-time limit, $t_2-t_1\gg a$ and $t_1\gg a$, time dependences are eliminated and the ratio in Eq.\eqref{eq:Ratiospinhalf} reduces to
  \begin{equation}
     R(t_2,t_1,{\bf p'},{\bf p},\Gamma,\mu) \xrightarrow[t_1\gg a]{t_2 -t_1 \gg a} \Pi({\bf p'},{\bf p};\Gamma;\mu).
     \label{eq:RtoPi}
  \end{equation}
One can extract the Sachs form factors by choosing the appropriate combinations of projection matrices $\Gamma$ and the Lorentz index $\mu$:
  \begin{equation}
      \Pi({\bf p'},{\bf p};\Gamma_4;\mu=4) = \sqrt{\frac{(E_{{\cal B}}+m_{{\cal B}})}{2E_{\cal B}}} G_E(q^2), 
     \label{eq:gE}
  \end{equation}
  \begin{equation}
     \Pi({\bf p'},{\bf p};\Gamma_j;\mu=i) =\sqrt{\frac{1}{2E_{\cal B}(E_{{\cal B}}+m_{{\cal B}})}} \epsilon_{ijk} q_k G_M(q^2). 
     \label{eq:gM}
  \end{equation}
Here $G_E$ gives the electric charge of the baryon at zero transferred momentum. Similarly, the magnetic moment can be obtained by extrapolating the $G_M$ to zero transferred momentum.

Simulations have been run on $32^3 \times 64$ lattices with $2+1$ flavors of dynamical quarks that have been generated by the PACS-CS collaboration~\cite{PACS-CS:2008bkb} with the non-perturbatively O(a)-improved Wilson quark action and the Iwasaki gauge action. Simulations are carried out with near physical $u$, $d$ sea quarks of hopping parameter $\kappa^{u,d}_{\text{sea}}= 0.13781$. Hopping parameter for the sea $s$ quark is fixed to $\kappa^{s}_{\text{sea}} = 0.13640$. Features of the lattice setup are explained in detail in Ref.~\cite{Bahtiyar:2016dom}.

We use the Clover action for the $u$, $d$, and $s$ valence quarks. Hopping parameter of the light-quarks is taken equal to that of the light sea quark $\kappa^{u,d}_{\text{val}}= \kappa^{u,d}_{\text{sea}} =0.13781$  which corresponds to a pion mass of approximately $156$~MeV~\cite{PACS-CS:2008bkb}. As for the strange quark, it is reported that the hopping parameter overestimates the experimental value by $6\%$~\cite{Bahtiyar:2018vub}. Therefore hopping parameter of the strange quark is taken as $\kappa^s_{\text{val}} = 0.13665$, as suggested in Refs.~\cite{Menadue:2011pd,Mohler:2013rwa}.

We utilize two different fermion actions for the charm quark. Firstly, we apply a Clover action in the form used by Fermilab and MILC Collaborations~\cite{El-Khadra:1996wdx,Burch:2009az,FermilabLattice:2010rur}. We use the value of the hopping parameter $\kappa^c_{\text{Clover}}=0.1246$ as determined in our previous work~\cite{Can:2013tna}. The second fermion action we implement is the relativistic heavy-quark action (Tsukuba) proposed by Aoki et al.~\cite{Aoki:2001ra}. This action is designed to reduce the leading cutoff effects, which can be removed by tuning the action's parameters non-perturbatively. As a result, only $\mathcal{O}((a \Lambda_{QCD})^2)$ discretization errors remain.  The hopping parameter is taken as $\kappa^c_{\text{Tsukuba}}=0.10954007$. The tuning of the action's parameters is explained in detail in Ref.~\cite{Bahtiyar:2018vub}.

We insert positive and negative momenta in all spatial directions and make a simultaneous fit over all data to increase statistics. We also calculate current insertion along all spatial directions. Data are binned at the bin size of $20$ in order to account for autocorrelations. The source-sink time separation is fixed to $12$ lattice units to prevent the excited state contamination~\cite{Can:2013tna,Bahtiyar:2018vub}. We also use multiple source-sink pairs, shifting them $12$ lattice units along the temporal direction. The point-split lattice vector current is employed
  \begin{equation}
		 J_\mu = \frac{1}{2} [\overline{q}(x+\mu) U_\mu^{\dagger}(1+\gamma_\mu)q(x)-\overline{q}(x)U_\mu(1-\gamma_\mu)q(x+\mu)],
		 \label{eq:pointsplit}
  \end{equation}
which is conserved by Wilson fermions and therefore does not need any renormalization for Clover action. The point-split lattice vector current is also conserved for the $G_E$ for the Tsukuba action. However, it is not conserved for the spatial components of the Tsukuba action; therefore, there needs to be a renormalization coefficient for the magnetic form factor. The lack of the renormalization coefficient in this work can be treated as a discretization error that arises due to the anisotropy present in the action; this concern needs to be addressed in future works. Momentum is inserted up to $9$ lattice units and averaged over equivalent momenta. All statistical errors are estimated by the single-elimination jackknife analysis. The wall-source/sink method~\cite{Can:2012tx} is employed; therefore, one can simultaneously extract all spin, momentum, and projection components. The wall source/sink is a gauge-dependent object; thus, the gauge is fixed to Coulomb.

To confirm that the ground baryon state is isolated from the excited-state contaminations, we performed further analysis as in Ref.~\cite{Can:2013tna} and fitted the ratio in Eq.~\eqref{eq:Ratiospinhalf} to a phenomenological form 
\begin{equation}\label{excfit}
	R(t_2,t_1)=G_{E,M}+b_1\,e^{-\Delta t_1}+b_2\,e^{-\Delta (t_2-t_1)},
\end{equation}
where $\Delta$ is the energy gap between the ground and the excited state. This method has been demonstrated to be useful in systematically analyzing excited-state contamination for nucleon form factors~\cite{Bhattacharya:2013ehc}. However, a challenge in applying this approach to doubly charmed baryons is that the energy gaps have not yet been determined. As a result, we treat $\Delta$ as a free parameter along with $b_1$ and $b_2$, which introduces greater uncertainty into $G_{E,M}$. Additionally, we note that the asymmetric smearing of the source and sink introduces a further complication, with $b_1\neq b_2$.

\begin{figure}[t]
    \centering
      \includegraphics[width=0.70\textwidth]{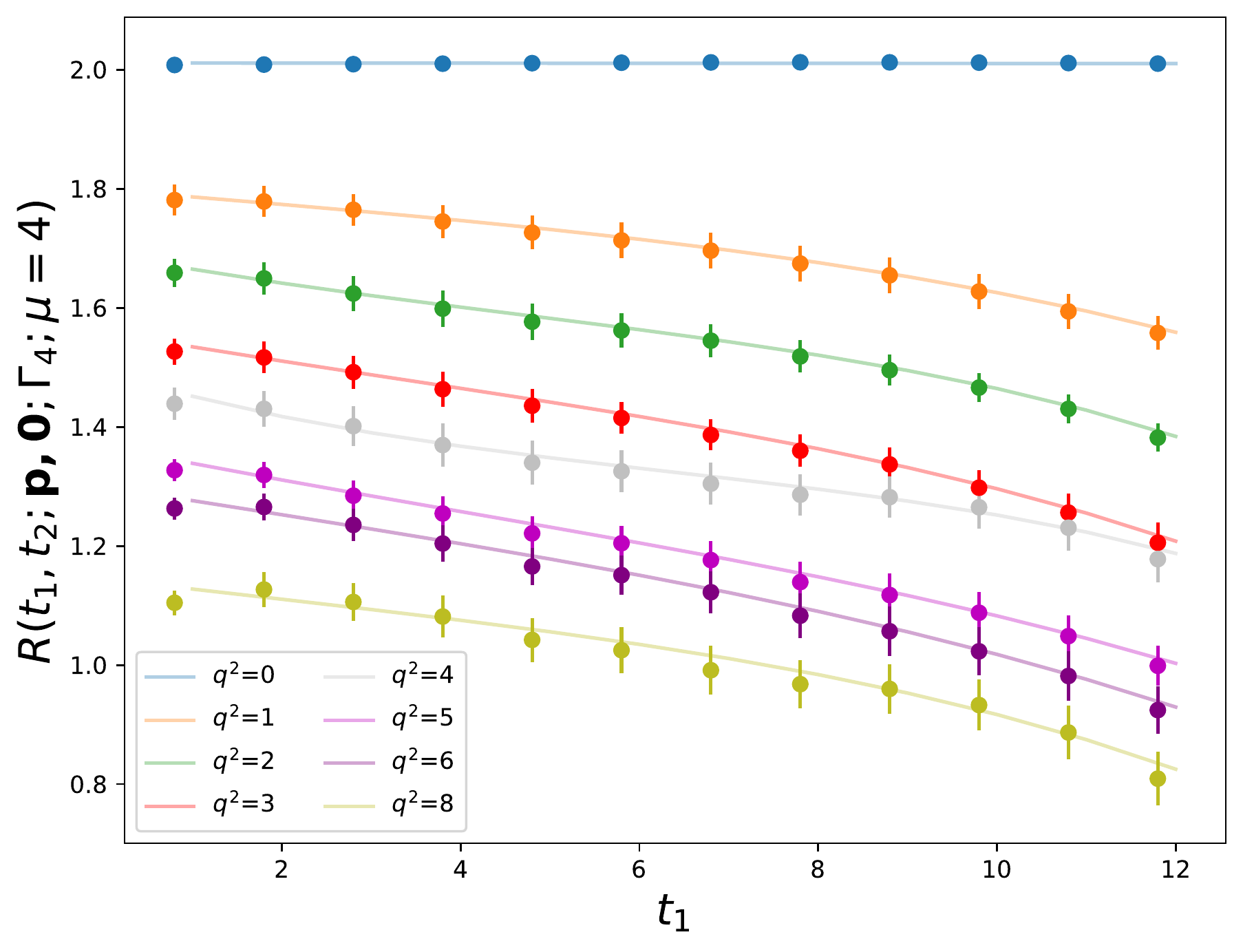}
    \caption{Two state fit results of $\Xi_{cc}^{++}$}
    \label{fig:twostate}
  \end{figure}

The two-state fit method successfully reproduces the data, as depicted in \Cref{fig:twostate}. The parameter values for the ratio of electric form factors of $\Xi_{cc}$ at all momentum transfers are presented in \Cref{tab:GEfits}. It is worth noting that the statistical error in the energy gap $\Delta$ is substantial. Therefore, we refrain from interpreting $\Delta$ as a physical energy gap. The statistical uncertainties in the fit parameters are also considerable, making it challenging to determine the magnetic form factor accurately. Nonetheless, the fit values of electric form factors fall within the margin of error, and thus we employ the approach with the lowest error for further analysis.


\begin{table}[ht]
\centering
\caption{The parameter values in the case of electric form factors of $\Xi_{cc}^{++}$ along with plateau fit results for all momentum transfers.}
\label{tab:GEfits}
	\setlength{\extrarowheight}{5pt}
	\begin{tabular*}{\textwidth}{@{\extracolsep{\fill}}l|l||l|l|l|l}
	\hline \hline
$Q^2$\,$[\frac{2\pi}{a\,N_s}]$ &$G_E$ plateau fit &$G_E$ & $\Delta$ & $b_1$ &$b_2$ \\
\hline
\hline
0     & 2.001 (12)                    & 2.010 (12)          & 0.100 (100)               & 0.010 (10)           & -0.001 (1)           \\ \hline
1     & 1.732 (28)                    & 1.792 (68)          & 0.154 (57)              & 0.036 (39)           & -0.282 (76)          \\ \hline
2     & 1.580 (29)                    & 1.589 (57)          & 0.220 (67)              & 0.096(29)           & -0.265 (51)          \\ \hline
3     & 1.436 (26)                    & 1.478 (100)         & 0.168 (90)              & 0.102 (32)           & -0.338 (111)         \\ \hline
4     & 1.361 (35)                    & 1.324 (51)          & 0.267 (117)             & 0.135 (40)           & -0.188 (68)          \\ \hline
5     & 1.244 (27)                    & 1.254 (139)         & 0.143 (116)             & 0.143 (58)           & -0.325 (168)         \\ \hline
6     & 1.194 (29)                    & 1.268 (79)          & 0.126 (58)               & 0.096 (56)           & -0.415 (100)         \\ \hline
8     & 1.068 (35)                    & 1.121 (185)         & 0.175 (132)             & 0.054 (70)           & -0.357 (152)       \\ \hline
\end{tabular*}
\end{table}

The finite-size effects should be negligible when $m_\pi L \geq 4$. On the other hand, the ensemble that we used in the analysis yields $m_\pi L = 2.3$, which is below this bound. However, it is confirmed that the finite-size effects on this particular setup are under control for physical quantities related to strange and charmed baryons~\cite{Can:2015exa}. 

Only connected diagrams are computed in this work. We performed our computations using a modified version of Chroma software system~\cite{Edwards:2004sx} on CPU clusters and with QUDA~\cite{Babich:2011np,Clark:2009wm} for propagator inversion on GPUs.

\section{Results and Discussion}
\label{sec:res}
We begin our work by extracting the masses using the two-point correlation function given in Eq.\eqref{eq:twopt}. If the sink operator is projected to zero momentum, the two-point correlation functions reduce to
\begin{equation}
\langle F^{{\cal B}\,{\cal B}}(t;{\bf p};\Gamma_4)\rangle  \simeq Z_{\cal B}(\mathbf{p}) \bar{Z}_{\cal B}(\mathbf{p}) e^{-E_{\cal B}(\mathbf{p}) t} (1 + \mathcal{O}(e^{-\Delta E t}) + \dots),
\end{equation}

where the mass of a baryon is encoded into the leading exponential and can be identified for the \textbf{p} = (0, 0, 0) case when the excited states are properly suppressed. We perform standard effective mass analysis using the forms given

\begin{align}
      \label{eq:effmass}
    m_{eff}\bigg(t+\frac{1}{2}\bigg) &= ln \frac{C(t)}{C(t+1)},\\
        \label{eq:expfit}
    C(t) &= Z e^{-m t}.
  \end{align}
When the ground state is dominant, the signal obtained from Eq.\eqref{eq:effmass} forms a plateau. To this end, we seek a plateau to estimate suitable fit ranges for the one-exponential fit function given in Eq.\eqref{eq:expfit}. Then we extract the ground state masses from the correlation functions. It is possible to take the contribution of the first excited state as a correction term to Eq.\eqref{eq:expfit} to pinpoint the ground state precisely; however, we find it to be an excessive treatment considering the precision and agreement of our results. We calculate the masses of the baryons created using the Clover and Tsukuba actions separately. Our results are given in \Cref{tab:mass}, together with the experimental values and those obtained by other lattice collaborations.

\begin{table}[ht]
\centering
\caption{Extracted $\Xi_{cc}$, and $\Omega_{cc}$ masses as well as those of other lattice collaborations and experimental values. The errors in this work are statistical only, while those quoted by other collaborations correspond to statistical and various systematical errors if given.}
\label{tab:mass}
	\setlength{\extrarowheight}{5pt}
	\begin{tabular*}{\textwidth}{@{\extracolsep{\fill}}lll}
	\hline \hline
	& $m_{\Xi_{cc}}$ [GeV] & $m_{\Omega_{cc}}$ [GeV]  \\ \hline
Tsukuba & 3.626 (30) &  3.720 (12)\\
Clover &  3.627 (32) & 3.726 (12) \\\hline
Experiment \cite{LHCb:2017iph} & 3.62140(72)(27)(14) & --- \\
PACS-CS \cite{PACS-CS:2013vie} & 3.603(22) & 3.704(17)\\
ETMC \cite{Alexandrou:2014sha} &  3.568(14)(19)(1) &  3.658(11)(16)(50)\\
Briceno et al. \cite{Briceno:2012wt} &  3.595(39)(20)(6) & 3.679(40)(17)(5) \\
Brown et al. \cite{Brown:2014ena}  & 3.610(23)(22) & 3.738(20)(20)\\
RQCD \cite{Perez-Rubio:2015zqb} & 3.610(21) &  3.713(16) \\
 	\hline \hline
 	\end{tabular*}
\end{table}

For the $\Xi_{cc}$ baryon, the mass results using the Clover and Tsukuba actions agree with those from the LHCb experiment. Therefore, it is seen that charm quark actions are correctly tuned. The $\Omega_{cc}$ baryon is not experimentally observed yet; nevertheless, our results agree with the results obtained by other lattice collaborations. As seen from \Cref{tab:mass}, the mass of the $\Omega_{cc}$ is expected to be between $3.650 - 3.750$ GeV.

\begin{sidewaysfigure*}[p]
	\centering
	\includegraphics[width=1\textwidth]{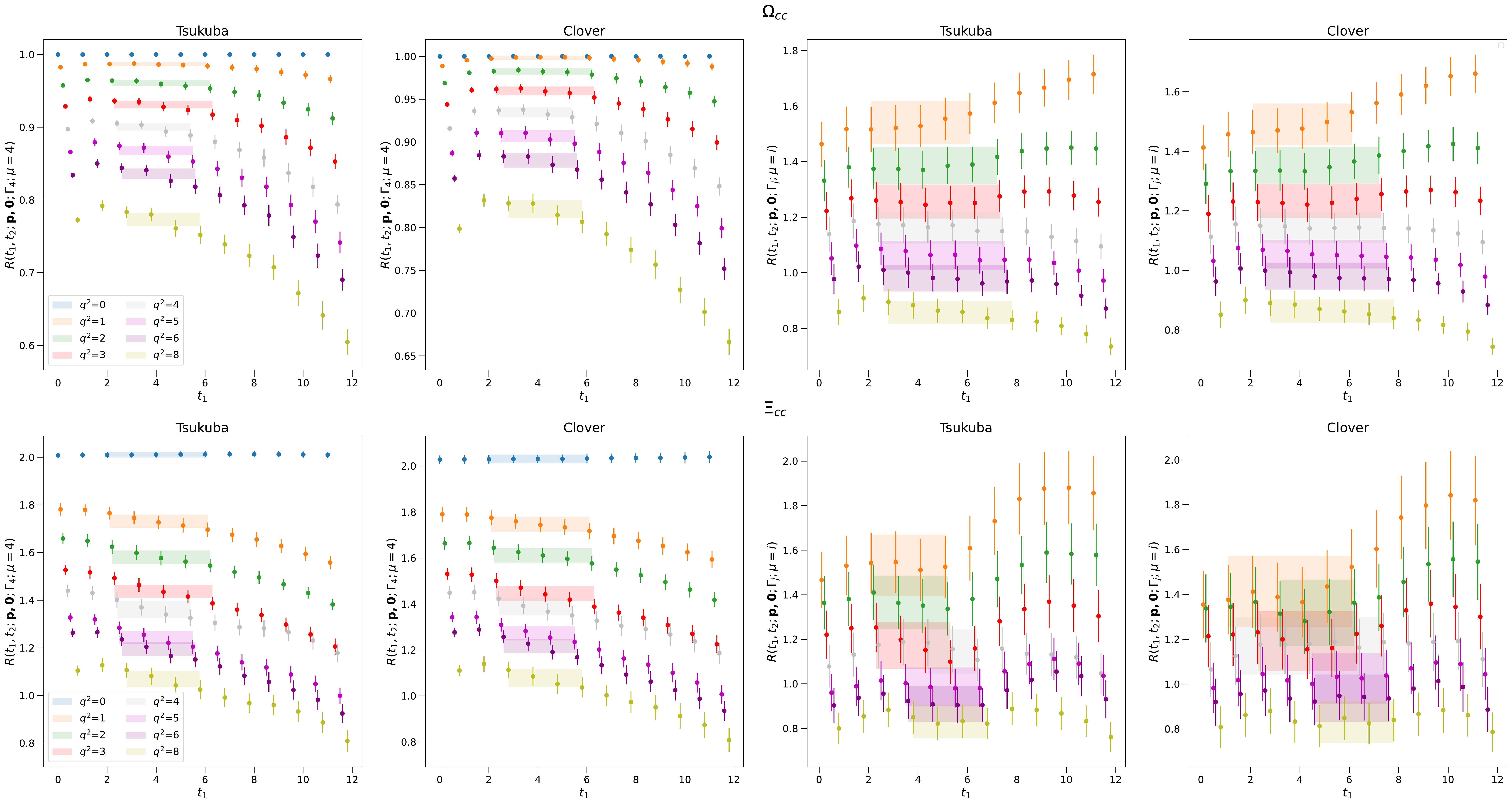}
	\caption{\label{el_plat} The ratio in Eq.~\eqref{eq:Ratiospinhalf} as function of the current insertion time, $t_1$, for the form factors of $\Xi_{cc}$ and $\Omega_{cc}$ baryons.}
	\label{fig:corrs}
\end{sidewaysfigure*}

We continue our work with calculating the ratios given in Eq.\eqref{eq:Ratiospinhalf}. We plot the correlators in \Cref{fig:corrs} as a function of current insertion time for each transferred three-momentum square and search for plateaus to exclude the excited state contamination. Ground state signals are found in the middle region between the source and sink points. We obtain fairly clean signals both for the $\Xi_{cc}$ and $\Omega_{cc}$. We use the p-value as a criterion in defining a plateau region~\cite{Can:2013zpa}. In each case, we look for plateau regions with at least three-time slices between the source and the sink and select the one with the greatest p-value. The weak coupling to the ground state and the associated excited-state contamination can explain the relatively strong late-time time dependence. As a result, regions closer to the smeared source are selected since they are projected to couple to the ground state with greater strength than the wall sink.

$G_E$ gives the electric charge of the baryon at zero transferred momentum, which can be computed directly with the formulation on the lattice. However, the formulation restricts to make a direct measurement of the magnetic form factor at zero momentum $G_M(0)$. To this end, we use the following dipole form to describe the $-q^2 \equiv Q^2$ dependence of the form factors
\begin{equation}
G_{E,M} (Q^2) = \dfrac{G_{E,M}(0)}{\left(1+Q^2/\Lambda_{E,M}^2\right)^2}.
\label{eq:dipoleform}
\end{equation}

\begin{figure}[t]
    \centering
      \includegraphics[width=0.99\textwidth]{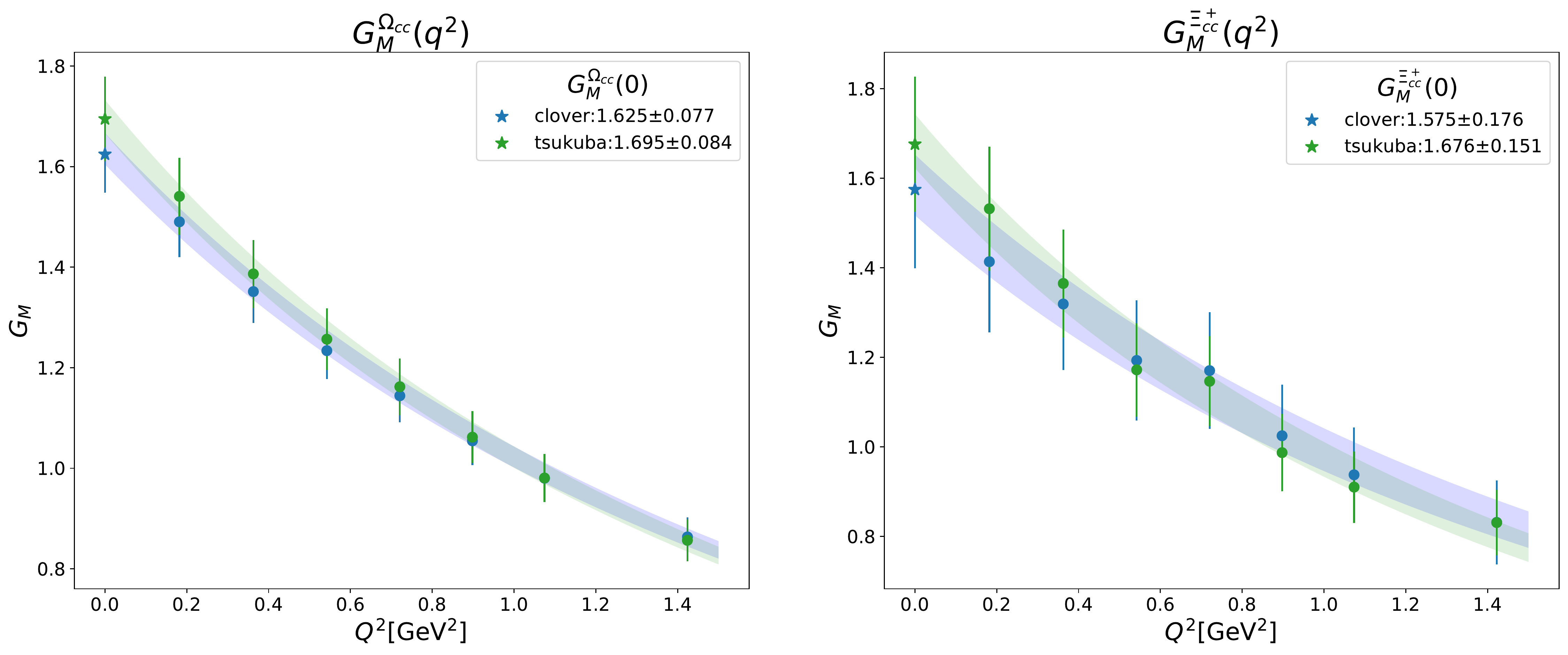}
    \caption{$Q^2$ dependence of the magnetic form factors of $\Xi_{cc}^{+}$ and $\Omega_{cc}$.}
    \label{fig:Mag}
  \end{figure}

\begin{figure}[t]
    \centering
      \includegraphics[width=0.99\textwidth]{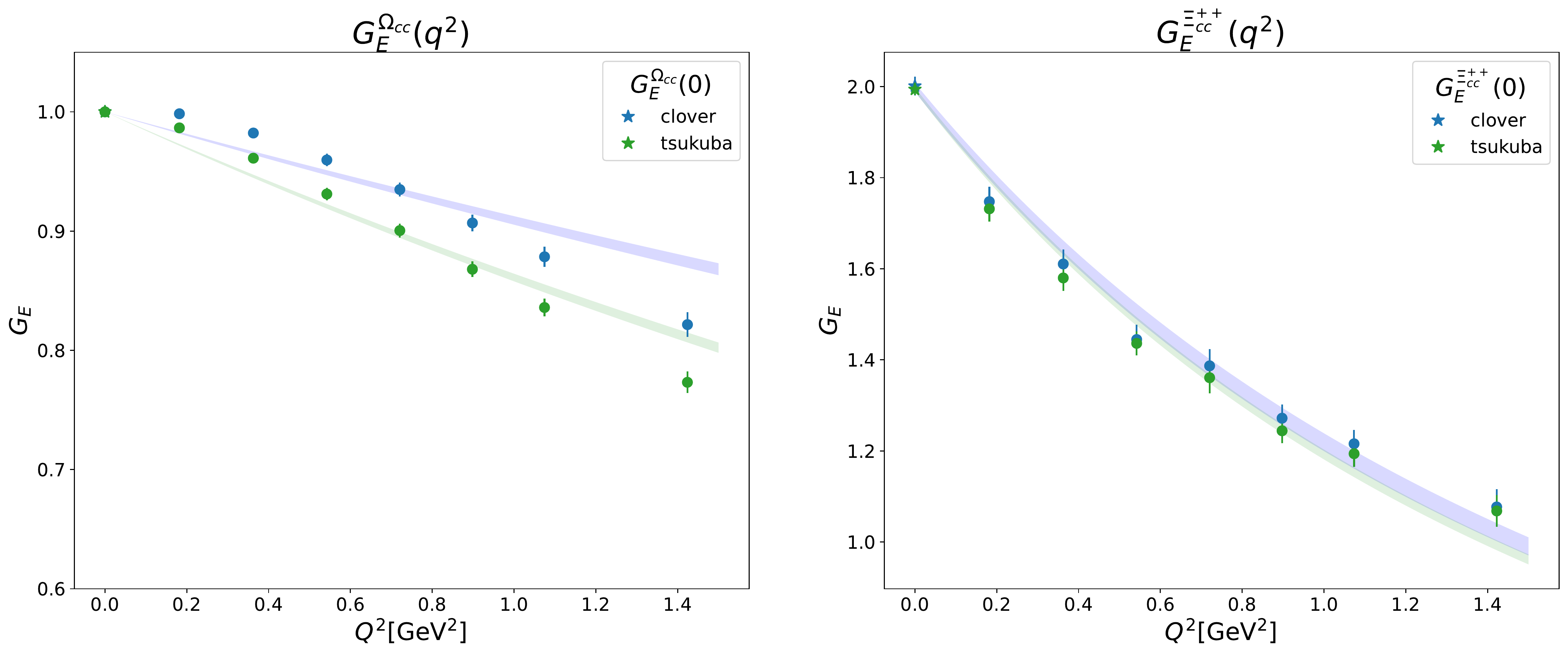}
    \caption{$Q^2$ dependence of the electric form factors of $\Xi_{cc}^{++}$ and $\Omega_{cc}$.}
	  \label{fig:EFF}
  \end{figure}

We examine the contribution of individual quark sectors to the magnetic properties to gain a deeper understanding of quark dynamics. The analysis is performed by coupling the electromagnetic field only to either the $u/s$ or the $c$ quark. The baryon form factors are calculated from individual quark contributions using Eq.\eqref{eq:magSquarkcont} by 
\begin{equation}
		G_{M} (Q^2) = \frac{2}{3} G_{M}^c (Q^2) + C_\ell G_{M}^\ell (Q^2),
		\label{eq:individual}
\end{equation}
where $c_\ell=-1/3$ for the $d$, $s$ quark, and $C_\ell=2/3$ for the $u$ quark. We combine the individual quark contributions using Eq.\eqref{eq:individual} for each momentum transfer $Q^2$ and extrapolate the combined form factor values to $Q^2=0$. The light and $c$ quark contributions for the $\Xi_{cc}$ have opposite signs, for $\Xi_{cc}^{+}$ $d$ and $c$ quark contributions are multiplied with electric charges of the opposite sign and added constructively. In contrast, contributions from the $u$ and $c$ quarks cancel each other out, resulting in noisy data for $\Xi_{cc}^{++}$. Therefore severe errors reported in $\Xi_{cc}^{++}$ form factors suffer from the poor signal-to-noise ratio and the limited number of gauge configurations. The reported results for $\Xi_{cc}^{++}$ still might be a useful constraint for comparison with quark models. Consequently, in this work, we can only focus on $\Xi_{cc}^{+}$ and $\Omega_{cc}$  magnetic form factors.~\Cref{fig:Mag,fig:EFF} displays the electric and magnetic form factors of $\Xi_{cc}$ and $\Omega_{cc}$ as functions of $Q^2$. Our results for the magnetic form factors are given in~\Cref{tab:moment}. As can be seen from the \Cref{fig:Mag}, the dipole form describes the lattice data successfully for most of the cases. The central values of the magnetic form factors calculated using the Tsukuba action are approximately $5\%$ greater than those of the Clover action. A similar pattern has already been observed in our spin-$3/2 \to$ spin-$1/2$ radiative transition work~\cite{Bahtiyar:2018vub}.  It is important to note that the observed $5\%$ discrepancy in the magnetic form factors may be attributed to the lack of conservation of the point-split lattice vector current in Eq.~\eqref{eq:pointsplit} for its spatial components. This discrepancy between the results obtained using the Clover and Tsukuba actions could potentially be mitigated by appropriately determining the current for both its temporal and spatial components and subsequently reevaluating the calculations.

On the other hand, it is noteworthy that the electric form factors of the $\Omega_{cc}$ baryon exhibit behavior that suggests a more substantial alignment with a linear trend, as evidenced by previous studies~\cite{Can:2013tna}. In our current investigation, we have fine-tuned the value of the strange quark ($\kappa^s_{\text{val}} = 0.13665$), resulting in the electric form factors of the $\Omega_{cc}$ baryon displaying a closer adherence to a linear functional form. This observation raises legitimate concerns regarding the reliability of the dipole form, as described by Eq.~\eqref{eq:dipoleform}, which we employed for all form factors.

To quantitatively evaluate the quality of our fits, we have calculated the $\chi^2$/dof values for the fits of the electric form factors associated with the Tsukuba and Clover actions. The obtained $\chi^2$/dof for Tsukuba action is 2.966, while for the Clover action, it is 1.732. It is important to note that these $\chi^2$/dof results provide an indication of the overall agreement between our fitting model and the lattice data. A value close to 1 signifies a good fit, suggesting that the model adequately captures the underlying physics. Conversely, a significantly larger value indicates a poor fit, suggesting that the model needs to capture the complexities present in the data. Consequently, we suspect that the poor fitting, as indicated by the relatively high $\chi^2$/dof values, may be the reason for the extracted results of $\langle r^2_E \rangle$ for $\Omega_{cc}$ close to zero.

Considering the relatively higher $\chi^2$/dof values obtained in our fits, exploring alternative approaches to parameterize the form factors more effectively is prudent. One such approach is utilizing the z-expansion technique, which offers greater flexibility in capturing the intricate dynamics and non-linearities inherent in the form factors. The z-expansion method can achieve a more comprehensive and robust analysis, leading to more reliable and accurate results.

In light of these considerations, we recognize the limitations associated with the dipole form utilized in our study. We recommend future investigations to explore alternative parameterizations, such as the z-expansion, to improve the accuracy and reliability of the results. By employing a more adaptable functional form, one can better account for the behavior of the electric form factors of the $\Omega_{cc}$ baryon, ensuring a more comprehensive understanding of its properties.

One can extract the charge radii of the baryons from the slope of the form factor at $Q^2=0$,
\begin{equation}
	\langle r_{E,M}^2 \rangle=-\frac{6}{G_\text{E,M}(0)} \frac{d}{dQ^2}G_{E,M}(Q^2)\bigg|_{Q^2=0}.
\end{equation}
To evaluate the charge radii with the above formula, we will assume the dipole form given in Eq.\eqref{eq:dipoleform} for the form factors leading to the relation,
\begin{equation}\label{eq:emfitform}
	\langle r_{E,M}^2 \rangle=\frac{12}{\Lambda_{E,M}^2}.
\end{equation}

Then the charge radii can be directly calculated using the values of the fit parameters obtained from the dipole fit to the form factor data. Our results for the electromagnetic charge radii are given in~\Cref{tab:moment}. One can compare the baryons with the same electric charge to figure out the effects of internal dynamics. It is seen in~\Cref{tab:moment} that electric charge radii of $\Xi_{cc}^+$ and $\Omega_{cc}$ baryons are very close to each other, and much smaller compared to proton's $\langle r_E^2 \rangle^p = 0.707$ fm$^2$~\cite{10.1093/ptep/ptaa104}. When comparing $\Xi_{cc}^+$ and $\Omega_{cc}$, the only difference is that a light-quark is changed to a strange quark, albeit this change shows that the strange quark has minimal effect on the electric charge radius. $\Xi_{cc}^{++}$ has the largest charge radius, as it has two units of electric charge. When we examine the contribution of individual quarks, it is apparent that the light-quark contributions are greater than the charm-quarks, and the main difference comes from the electric charges in the baryon. Overall the results agree with our previous findings~\cite{Can:2013tna}, and it can be summarized from a quark model point of view as the heavy $c$ quark core acts to shift the center of mass towards itself, reducing the size of the baryon. The results calculated using different actions differ from each other by $5$ to $10$\%, and the electric charge radii calculated with the Tsukuba action are greater than those from the Clover action.

Magnetic charge radii of $\Xi_{cc}^+$ and $\Omega_{cc}$ have similar behavior to their electric charge radii and are comparable to each other. The light-quark contribution is observed to be greater than the charm-quark contribution. Note that the magnetic form factor of $\Xi_{cc}^{++}$ is too noisy to make a proper comment due to a poor signal-to-noise ratio.

In order to calculate the magnetic moment,  one needs to extract the $G_M$ at zero transferred momentum using the dipole form in Eq.\eqref{eq:dipoleform}, then magnetic moment can be calculated using the equation below

\begin{equation}
\mu_{\cal B} = G_M(0) \Big(\frac{e}{2m_{\cal B}}\Big)= G_M(0) \Big(\frac{m_N}{m_{\cal B}}\Big)\mu_N,
\label{eq:magneticmoment}
\end{equation}
where $m_N$ is the physical nucleon mass and $m_{\cal B}$ is the baryon mass obtained in this work given in~\Cref{tab:mass}. Our results for the magnetic moments are given in~\Cref{tab:moment}.

The Tsukuba and Clover nomenclature in~\Cref{tab:moment} applies only to charm quarks. All light and strange quarks in this work are calculated using the Clover action. Since all the quarks in the baryon are bound to each other, the light quark results of the baryon calculated by the Tsukuba action of the charm quark propagator differ from the baryon calculated with the Clover action.

The magnetic moment of $\Xi_{cc}^+$ and $\Omega_{cc}$ are similar. The light-quark contribution is found to be negative, and the absolute value of the light-quark contribution is greater than the charm-quark contribution. The signs of the magnetic moments disclose the interaction of the spins of the quarks. The opposite signs of the light- and heavy-quark magnetic moments mainly indicate that their spins are anti-aligned. As given in~\Cref{tab:moment}, the light- and heavy-quark magnetic moments are of opposite signs, so by simple deduction, the charm quarks are paired in a spin-1 state with their spins aligned, which leads to a significant charm quark contribution to the total spin and magnetic moment~\cite{Can:2013tna}.
\begin{table}[ht]
\centering
\caption{The form factor values extrapolated to $Q^2=0$, together with the magnetic moments in units of nuclear magneton. The Tsukuba and Clover nomenclature applies only to charm quarks. All light and strange quarks in this work are calculated using the Clover action. The results are normalized to unit contribution. Note: $\langle r^2_E \rangle$ for $\Omega_{cc}$ are also extracted using dipole form given in Eq.~\eqref{eq:dipoleform}}
\label{tab:moment}
	\setlength{\extrarowheight}{5pt}
	\begin{tabular*}{\textwidth}{@{\extracolsep{\fill}}l|cc|cc|cc|cc}
	\hline \hline
 & \multicolumn{2}{c|}{$G_M(0)$}  & \multicolumn{2}{c|}{Magnetic moment~[$\mu_N$]}  & \multicolumn{2}{c|}{$\langle r^2_E \rangle$ [fm$^2$]}& \multicolumn{2}{c}{$\langle r^2_M \rangle$ [fm$^2$]}   \\ 
 & Tsukuba & Clover & Tsukuba & Clover& Tsukuba & Clover& Tsukuba & Clover\\\hline
 $ \Xi_{cc}^\ell$  & -1.218 (142) & -1.119 (134) & -0.315 (37) & -0.290 (35)   & 0.443 (37) & 0.468 (51) & 0.353 (30) & 0.299 (27) \\
 $ \Xi_{cc}^c$  & 1.681 (147) & 1.636 (183) & 0.435 (38)& 0.423 (48) & 0.090 (5) & 0.084 (5) & 0.080 (2) & 0.070 (3)  \\
 $ \Xi_{cc}^{+}$  & 1.676 (151) & 1.575 (176) & 0.433 (39) & 0.407 (45) & 0.024 (8) & 0.016 (8) & 0.148 (3) & 0.123 (4)\\
 $ \Xi_{cc}^{++}$  & 0.301 (200) & 0.320 (191) & 0.080 (52) & 0.089 (45) & 0.137 (6) & 0.132 (8) & 0.288 (210) & 0.250 (200) \\ \hline
 $ \Omega_{cc}^s$ & -1.659 (83) & -1.653 (78)  & -0.418 (21) & -0.416 (20) & 0.296 (9) & 0.281 (8) & 0.389 (2) & 0.382 (3)\\
 $ \Omega_{cc}^c$  & 1.824 (90)& 1.727 (80)  & 0.460 (23) & 0.434 (21) & 0.091 (2)& 0.087 (2) &  0.091 (3) & 0.079 (2)\\
 $ \Omega_{cc}$  & 1.695 (84)& 1.625 (77) & 0.430 (19) & 0.416 (18) & 0.036 (3) & 0.022 (3)  & 0.131 (5) & 0.127 (5) \\ 
 	\hline \hline
 	\end{tabular*}
\end{table}


The ensemble we used in this work has almost physical light-quark masses at $m_\pi \approx 156$~MeV; therefore, extrapolation to the chiral limit is not strictly necessary. Nevertheless, we combine our findings of Clover action with our previous results given in Ref.~\cite{Can:2013tna} to investigate pion-mass dependence as we approach the physical point. To obtain the values of the observables at the chiral point, we perform fits that are linear and quadratic in $m_\pi^2$:
\begin{align}
f_{lin} &= a m_\pi^2 + b\\
f_{quad} &= c m_\pi^4 + d m_\pi^2 + e
\end{align}
where a,b,c,d,e are the fit parameters. In order to keep consistency, we only make chiral extrapolation to observables of $\Xi_{cc}^+$ since the hopping parameter of strange valance quark differs from our previous work~\cite{Can:2013tna}.

  \begin{figure}[t]
    \centering
      \includegraphics[width=0.99\textwidth]{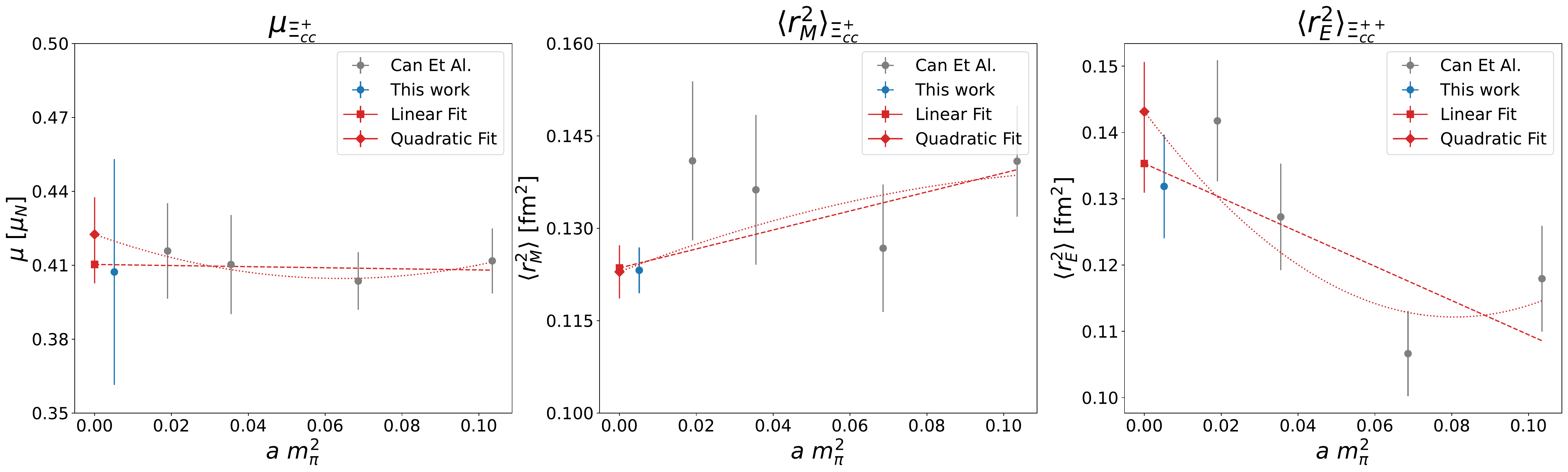}
    \caption{Chiral extrapolations of magnetic moment (left), and magnetic (middle) and electric (right) charge radii of $\Xi_{cc}^{+}$. Gray data points are taken from Ref.~\cite{Can:2013tna}, blue data points are our results calculated with Clover action, red diamonds and red squares represent the quadratic and linear fits respectively.}
\label{fig:chiral}
  \end{figure}

\Cref{fig:chiral} shows the chiral fits of $\Xi_{cc}^+$. On the left panel, we show the magnetic moment as a function of the square of the pion mass in lattice units. The magnetic moment appears to be more compatible with the linear form showing almost no quark mass dependence. The figure in the middle shows the chiral fits of the magnetic charge radius. It is seen that the quark mass dependence is described by linear form better. In the figure on the right, the electric radius is given. The errors agree with our previous findings, and there seems to be mild pion mass dependence.

It is not possible to make a similar extrapolation for $\Omega_{cc}$ as a different hopping parameter is used compared to Ref~\cite{Can:2013tna}. However, a quick qualitative analysis is applicable; values of the observables increase around $10-15\%$ due to the strange-quark retuning~\cite{Bahtiyar:2018vub}. If we take this into account and make a chiral extrapolation, we observe similar quark mass dependence behavior to \Cref{fig:chiral}. These findings suggest that observables of doubly charmed baryons are mildly dependent on the sea-quark mass.

As a result, one can conclude that the electromagnetic observables we have calculated near the physical point using the Tsukuba action can be taken as our final values. These conclusions corroborate the findings of our previous works~\cite{Bahtiyar:2020uuj,Bahtiyar:2015sga,Bahtiyar:2018vub,Bahtiyar:2016dom}.

\begin{table}[ht]
\centering
\caption{Comparison of our results with various other models. All values are given in nuclear magnetons.}
\label{tab:comparison}
	\setlength{\extrarowheight}{5pt}
	\begin{tabular*}{0.5\textwidth}{@{\extracolsep{\fill}}l|l|l}
	\hline \hline
& $\mu_{\Omega_{cc}^{+}}~$[$\mu_N$] & $\mu_{\Xi_{cc}^+}~$[$\mu_N$] \\ \hline
This Work & 0.430 (19) & 0.433 (39) \\
Lattice QCD~\cite{Can:2013tna}& 0.413 (24) & 0.425 (29)  \\
QCD S.R.~\cite{Ozdem:2018uue} & 0.39 (9) & 0.43 (9)\\
R. Quark Model~\cite{Julia-Diaz:2004yqv} & 0.72  & 0.86\\
N. R. Quark Model~I~\cite{Patel_2008}&  0.785 & 0.860 \\
N. R. Quark Model~II~\cite{Albertus:2006ya}&  $0.635^{+0.012}_{-0.015}$ & $0.785^{+0.050}_{-0.030}$ \\
R. T. Quark Model~\cite{Faessler:2006ft} & 0.67 & 0.74 \\
C.C. Quark Model~\cite{Sharma:2010vv} & 0.697 & 0.84\\
MIT Bag model~I~\cite{Bernotas:2012nz} & 0.668  & 0.722 \\
MIT Bag model~II~\cite{Simonis:2018rld}& 0.645  & 0.719 \\
MIT Bag model~III~\cite{Zhang:2021yul}  & 0.86 &  0.91\\
EOMS BHCPT~I~\cite{HillerBlin:2018gjw} & 0.40 (3)  & 0.37 (2)\\
EOMS BHCPT~II~\cite{Liu:2018euh} & 0.397 (15) & 0.392 (13)\\
HB ChPT~\cite{Li:2020uok}& 0.41 &  0.62  \\
 	\hline \hline
 	\end{tabular*}
\end{table}

Electromagnetic properties of doubly charmed baryons have been previously studied in quark models~\cite{Mutuk:2021epz,Julia-Diaz:2004yqv,Patel_2008,Albertus:2006ya,Faessler:2006ft,Jena:1986xs,Sharma:2010vv}, MIT bag model~\cite{Bernotas:2012nz,Simonis:2018rld,Zhang:2021yul}, chiral perturbation theory~\cite{HillerBlin:2018gjw,Liu:2018euh,Li:2017cfz,Li:2020uok}, QCD sum rules~\cite{Aliyev:2022rrf,Ozdem:2018uue} and Lattice QCD~\cite{Can:2013zpa,Can:2013tna}. Our final results for the magnetic moments are given in~\Cref{tab:comparison} along with a comparison to the literature. The signs of the magnetic moments are correctly determined. However, there is a discrepancy among the results. The moments seem to be underestimated with respect to quark models~\cite{Julia-Diaz:2004yqv,Patel_2008,Albertus:2006ya,Faessler:2006ft,Sharma:2010vv}, bag models~\cite{Bernotas:2012nz,Simonis:2018rld,Zhang:2021yul}, however, our findings are in agreement with those obtained from QCD sum rules~\cite{Ozdem:2018uue} and extended on-mass-shell chiral perturbation theory~\cite{HillerBlin:2018gjw,Liu:2018euh}. These findings are also in agreement with our earlier observations~\cite{Can:2013zpa,Can:2013tna}.

\section{Conclusion}
\label{sec:con}
The present work has aimed to examine the electromagnetic properties of doubly charmed baryons from $2+1$-flavor near physical light-quark masses simulations on $32^3 \times 64$ lattice. We have extracted the magnetic moments, and the electric and magnetic charge radii of $\Xi_{cc}$ and $\Omega_{cc}$. We have also determined individual quark contributions to the observables, which give valuable insight into the dynamics of the quarks having masses at different scales. The results of this investigation show that the doubly charmed baryons are compact in comparison to light baryons. The analysis of quark sector contributions to the charge radii has shown that light quark distributions are larger, and the heavy quark decreases the size of the baryon. The light- and heavy-quark contributions are opposite signs, which indicate that the charm quarks are paired in a spin-1 state. The magnetic moments seem to be underestimated compared to quark and bag models, but our findings agree with those obtained from QCD sum rules and extended on-mass-shell chiral perturbation theory. Lastly, this study has found that the observables of doubly charmed baryons are mildly dependent on the sea-quark mass.  

\section*{Acknowledgments}
The unquenched gauge configurations employed in our analysis were generated by PACS-CS collaboration~\cite{PACS-CS:2008bkb}. We used a modified version of Chroma software system~\cite{Edwards:2004sx} along with QUDA~\cite{Babich:2011np,Clark:2009wm}. The publicly available configurations are downloaded via the ILDG/JLDG network~\cite{Amagasa:2015zwb}. The author thanks G. Erkol, and K. U. Can from the TRJQCD Collaboration for valuable discussions and their comments on the manuscript. The author thanks E. Yüksel for reading early drafts of the manuscript and comments and O. Kaşıkçı for discussion on Ward Identities.
\bibliography{xicc.bbl}

\end{document}